\documentclass[12pt]{article}
\usepackage{times}
\usepackage{amsfonts}
\usepackage{amssymb}
\usepackage{amsmath}
\usepackage{latexsym}
\usepackage{amsthm}
\usepackage{fullpage}

\newcommand{\op}[1]{\operatorname{#1}}
\renewcommand{\t}{{\scriptscriptstyle\mathsf{T}}}
\newcommand{\ip}[2]{\left\langle #1 , #2\right\rangle}
\newcommand{\norm}[1]{\left\| #1 \right\|}
\newcommand{\abs}[1]{\left| #1 \right|}
\newcommand{\lin}[1]{\mathrm{L}\left(#1\right)}

\newcommand{\trans}[1]{\mathrm{T}\left(#1\right)}
\def \tr{\op{tr}}

\newenvironment{mylist}[1]{\begin{list}{}{
	\setlength{\leftmargin}{#1}
	\setlength{\rightmargin}{0mm}
	\setlength{\labelsep}{2mm}
	\setlength{\labelwidth}{8mm}
	\setlength{\itemsep}{0mm}}}
	{\end{list}}

\newtheorem{theorem}{Theorem}
\newtheorem{lemma}[theorem]{Lemma}
\newtheorem{cor}[theorem]{Corollary}
\newtheorem{prop}[theorem]{Proposition}
\theoremstyle{definition}

\begin{document}

\title{\Large\bf Bipartite subspaces having no LOCC-distinguishable bases}

\author{
John Watrous\\
Institute for Quantum Information Science and\\
Department of Computer Science\\
University of Calgary\\
Calgary, Alberta, Canada
}

\date{March 18, 2005}

\maketitle

\begin{abstract}
It is proved that there exist subspaces of bipartite tensor product spaces
that have no orthonormal bases that can be perfectly distinguished by means of
LOCC protocols.
A corollary of this fact is that there exist quantum channels having
sub-optimal classical capacity even when the receiver may communicate
classically with a third party that represents the channel's environment.
\end{abstract}


\section{Introduction}

One of the main focuses of the theory of quantum information in recent years
has been to understand the powers and limitations of {\em LOCC protocols}.
These are protocols wherein two or more physically separated parties 
possess the ability to perform arbitrary operations on local quantum systems
and to communicate with one another, but only classically.
The paradigm of LOCC, short for {\em local operations and classical
communication}, provides a setting in which to address basic questions
about the nature of entanglement and non-locality, generally viewed as
principal characteristics of quantum information.

One question along these lines that has received a great
deal of attention is that of {\em LOCC distinguishability} of sets of states.
In the two-party case, the two parties (Alice and Bob) share one of a known
orthogonal collection of pure states, and their goal is to determine which
of the states it is \cite{BennettD+99, BennettD+99a, ChenL03, Fan04, GhoshK+01,
GhoshK+04, HorodeckiS+03, Nathanson04, WalgateH02, WalgateS+00}.
In some cases it is possible for Alice and Bob to perform this task without
error and in some it is not.
For example, the fundamental result of Walgate, et~al.~\cite{WalgateS+00}
establishes that any two orthogonal pure states can be distinguished without
error.
On the other hand, large sets of maximally entangled states cannot; for
instance, if Alice and Bob's systems each correspond to $n$ dimensional
spaces, then it is impossible for them to perfectly distinguish $n+1$ or more
maximally entangled states \cite{Nathanson04}.
Other examples of sets of orthogonal states that cannot be perfectly
distinguished by LOCC protocols include those of \cite{BennettD+99}
and any set of states forming an unextendable product basis 
\cite{BennettD+99a}.
These examples demonstrate that entanglement is not an essential feature
of LOCC indistinguishable sets of states given that these sets contain only
product states.

This paper considers a related question, which is whether there exist
subspaces of bipartite tensor product spaces such that no orthonormal basis
of the subspace has the property that its elements can be perfectly
distinguished by means of an LOCC protocol.
Many examples of LOCC-indistinguishable sets fail to give an example of
such a subspace in that they span subspaces for which one can easily find
a perfectly distinguishable basis.
For example, the four Bell states are not perfectly distinguishable by any
LOCC protocol, but the space spanned by these states obviously does have a
perfectly distinguishable basis---the standard basis.
Indeed, {\em every} subspace of a tensor product space
$\mathcal{A}\otimes\mathcal{B}$ for which
$\op{dim}(\mathcal{A}) = \op{dim}(\mathcal{B}) = 2$ has a basis whose
elements can be perfectly distinguished by some LOCC protocol, and therefore
fails to have the property we are considering.
We prove, however, that if the dimension of both $\mathcal{A}$ and
$\mathcal{B}$ is at least three, then there do exist subspaces of 
$\mathcal{A}\otimes\mathcal{B}$ with the property that no basis of
the subspace is LOCC distinguishable.
In particular, it is proved that in the case 
$n = \op{dim}(\mathcal{A}) = \op{dim}(\mathcal{B})$ for $n\geq 3$,
the subspace of dimension $n^2 - 1$ that is orthogonal to the canonical
maximally entangled state (or any other fixed maximally entangled state)
has this property.

One motive for investigating this property is to identify quantum channels
having suboptimal classical corrected capacity with respect to the
definition of Hayden and King~\cite{HaydenK04}.
More specifically, Hayden and King considered the situation in which a sender
transmits classical information over a quantum channel to a receiver, who
has the added capability to measure the environment and use the result to
correct the channel's output.
This notion of correcting the output of a quantum channel by measuring
the environment was considered earlier by Gregoratti and
Werner \cite{GregorattiW04}, who focused primarily on the quantum capacity of
such channels.
Based on the result of Walgate, et al.~\cite{WalgateS+00}, Hayden and King
proved that the {\em classical corrected capacity} of any quantum channel
is at least one bit of information.
Many natural examples of channels can easily be seen to in fact have
{\em optimal} classical corrected capacity, meaning that the capacity
is $\log_2 n$ for $n$ the dimension of the input space, and no examples of
channels were previously proved to have less than optimal classical corrected
capacity.
The existence of subspaces having no LOCC distinguishable bases implies the
existence of such channels, even if the definition of Hayden and King is
extended to allow two-way communication between the receiver and the
environment.

The remainder of this paper is organized as follows.
Section~\ref{sec:preliminaries} discusses notation and background information,
Section~\ref{sec:indistinguishability} contains a proof of the main result
of the paper, which is that there exist subspaces of bipartite tensor product
spaces having no LOCC distinguishable bases, and Section~\ref{sec:channel}
discusses the implications of this result to classical corrected capacities
of quantum channels.
The paper concludes with a short list of open questions.


\section{Preliminaries}
\label{sec:preliminaries}

\subsection*{Basic notation}

This paper will use standard mathematical notation rather than Dirac notation
to represent vectors and linear mappings.
All vector spaces discussed in this paper are assumed to be finite dimensional
complex vector spaces.
The standard basis of a vector space $\mathcal{X}$ of the form
$\mathcal{X} = \mathbb{C}^n$ is $\{e_1,\ldots,e_n\}$, where $e_i$ is the
elementary unit vector defined by $e_i[j] = \delta_{ij}$.
The space of linear mappings from a space $\mathcal{Y}$ to a space
$\mathcal{X}$ is denoted $\lin{\mathcal{Y},\mathcal{X}}$, and we write
$\lin{\mathcal{X}}$ as shorthand for $\lin{\mathcal{X},\mathcal{X}}$ and
$\mathcal{X}^{\ast}$ as shorthand for $\lin{\mathcal{X},\mathbb{C}}$.
If $\mathcal{X} = \mathbb{C}^n$ and $\mathcal{Y} = \mathbb{C}^m$, then
elements of $\mathcal{X}$ are identified with $n$ dimensional column vectors,
elements of $\mathcal{X}^{\ast}$ are identified with $n$ dimensional row
vectors, and elements of $\lin{\mathcal{Y},\mathcal{X}}$ are identified with
$n\times m$ matrices in the typical way.
For $x\in\mathcal{X}$ we let $\overline{x} \in \mathcal{X}$ and
$x^{\t},x^{\ast}\in\mathcal{X}^{\ast}$ denote the entry-wise complex
conjugate, transpose, and conjugate transpose of $x$, and similar
for linear mappings; $\overline{X}\in\lin{\mathcal{Y},\mathcal{X}}$ and
$X^{\t},X^{\ast}\in\lin{\mathcal{X},\mathcal{Y}}$ denote the
entry-wise complex conjugate, transpose, and conjugate transpose of
$X\in\lin{\mathcal{Y},\mathcal{X}}$.
The usual inner products on $\mathcal{X}$ and
$\lin{\mathcal{Y},\mathcal{X}}$ are given by $\ip{x}{y} = x^{\ast}y$ and
$\ip{X}{Y} = \tr(X^{\ast}Y)$ for $x,y\in\mathcal{X}$ and 
$X,Y\in\lin{\mathcal{Y},\mathcal{X}}$.
The standard basis of the space $\lin{\mathcal{Y},\mathcal{X}}$ consists
of the mappings $E_{i,j} = e_i e_j^{\ast}$ for $1\leq i\leq n$ and
$1\leq j\leq m$.

The identity operator acting on a given space $\mathcal{X}$ is denoted
$I_{\mathcal{X}}$, or just as $I$ when $\mathcal{X}$ is implicit of
otherwise understood.
It is sometimes helpful to give different names to distinct but otherwise
identical spaces; in particular, we assume that $\mathcal{A} = \mathbb{C}^n$
and  $\mathcal{B} = \mathbb{C}^n$ are vector spaces referring to Alice's and
Bob's systems, respectively, throughout the paper.
We define $I_{\mathcal{B}, \mathcal{A}}\in\lin{\mathcal{B},\mathcal{A}}$ to
be the linear mapping that identifies vectors in $\mathcal{A}$ with vectors
in $\mathcal{B}$ by identifying the standard bases of these spaces.
Often this mapping is used implicitly.
For instance, if $a\in\mathcal{A}$ and
$b \in \mathcal{B}$ then $\ip{a}{b}$ is shorthand for
$\ip{a}{I_{\mathcal{B},\mathcal{A}}b}$, and when 
$X \in \lin{\mathcal{A},\mathcal{B}}$ we write $\tr (X)$ to mean
$\tr (I_{\mathcal{B},\mathcal{A}} X)$.

It is convenient when discussing bipartite quantum states to define a linear
bijection
\[
\op{vec}:\lin{\mathcal{Y},\mathcal{X}}\rightarrow\mathcal{X}\otimes\mathcal{Y}
\]
by the action $\op{vec}(E_{i,j}) = e_i\otimes e_j$ on standard basis elements,
extending by linearity.
It is simple to verify that for any choice of linear mappings $A$, $X$, and $B$
(for which the product $A X B$ is sensible), the equation
\[
(A\otimes B^{\t})\op{vec}(X) = \op{vec}(A X B)
\]
is satisfied.
For $\mathcal{A} = \mathbb{C}^n$ and $\mathcal{B} = \mathbb{C}^n$, the unit
vector
\[
\frac{1}{\sqrt{n}}\op{vec}(I_{\mathcal{B},\mathcal{A}})
= \frac{1}{\sqrt{n}}\sum_{i = 1}^n e_i\otimes e_i 
\in\mathcal{A}\otimes\mathcal{B}
\]
represents the canonical maximally entangled pure state in the
space $\mathcal{A}\otimes\mathcal{B}$.
Let $P\in\lin{\mathcal{A}\otimes\mathcal{B}}$ represent the projection
onto the space spanned by this vector,
\[
P = \frac{1}{n}
\op{vec}(I_{\mathcal{B},\mathcal{A}})
\op{vec}(I_{\mathcal{B},\mathcal{A}})^{\ast},
\]
and let $Q\in\lin{\mathcal{A}\otimes\mathcal{B}}$ denote the projection onto
the orthogonal complement of this space,
\[
Q = I_{\mathcal{A}\otimes\mathcal{B}} - P.
\]
Also let $\mathcal{P}$ and $\mathcal{Q}$ denote the subspaces of
$\mathcal{A}\otimes\mathcal{B}$ onto which $P$ and $Q$ project.


\subsection*{Separable measurements and perfect distinguishability}

There is no simple characterization known for the set of measurements that can
be realized by means of LOCC protocols.
For this reason it will simplify matters greatly for us to consider the set
of {\em separable measurements}, which does have a simple mathematical
characterization that we now discuss.

Let $\mathcal{A}$ and $\mathcal{B}$ be spaces corresponding to
two parties Alice and Bob.
A {\em separable measurement} on $\mathcal{A}\otimes\mathcal{B}$ with
possible outcomes $\{1,\ldots,N\}$ is a POVM described by a collection
\[
\{A_i\otimes B_i\,:\,i = 1,\ldots,N\}\subset
\lin{\mathcal{A}\otimes\mathcal{B}}.
\]
Similar to ordinary POVMs, $A_i$ and $B_i$ must be positive semidefinite
operators for each $i$, and must satisfy
\[
\sum_{i = 1}^N A_i\otimes B_i = I_{\mathcal{A}\otimes\mathcal{B}}.
\]
If we have that each of the operators $A_i$ and $B_i$ has rank equal to one,
we will say that the measurement is a {\em rank one separable measurement}.
Any measurement that can be realized by means of an LOCC protocol can be
described by a rank one separable measurement in the sense of the following
proposition.

\begin{prop}
Suppose that $\{M_k\,:\,k=1,\ldots,m\}$ is a POVM that describes the classical
output of a given LOCC protocol on $\mathcal{A}\otimes\mathcal{B}$.
Then there exists a rank one separable measurement
\[
\{a_i a_i^{\ast} \otimes b_i b_i^{\ast}\,:\,i = 1,\ldots,N\}
\]
on $\mathcal{A}\otimes\mathcal{B}$ together with a partition
$S_1 \cup \cdots \cup S_m = \{1,\ldots,N\}$,
$S_k\cap S_l = \varnothing$ for $k\not=l$, such that
\[
M_k = \sum_{i\in S_k} a_i a_i^{\ast} \otimes b_i b_i^{\ast}
\]
for $1\leq k \leq m$.
\end{prop}

\noindent
The fact that the classical output of any LOCC protocol can be described
by a separable measurement is well-known and the proof is routine.
It seems to have been first observed by Vedral and Plenio \cite{VedralP98} and
is discussed further in references \cite{BennettD+99, Rains97}.
By considering the spectral decomposition of its POVM elements, any separable
measurement can easily be further resolved to have rank one as claimed by
the proposition.
We note that the converse of the theorem is known to be false, as there exist
separable measurements that cannot be realized by LOCC
protocols \cite{BennettD+99}.

Suppose that $u_1,\ldots,u_m\in\mathcal{A}\otimes\mathcal{B}$ is a collection
of unit vectors.
A separable measurement $\{A_i\otimes B_i\,:\,i = 1,\ldots,N\}$ may be said to
{\em perfectly distinguish} this collection of vectors if there exists a
partition $S_1\cup \cdots \cup S_m = \{1,\ldots,N\}$, 
$S_k\cap S_l = \varnothing$ for $k\not=l$, such that
\[
u_k^{\ast} \left(\sum_{i\in S_l} A_i\otimes B_i\right) u_k = \delta_{kl}
\]
for $1\leq k,l\leq m$.

\begin{cor}
If Alice and Bob can perfectly distinguish the states $u_1,\ldots,u_m$
by means of an LOCC protocol, then there exists a rank one separable
measurement 
\[
\{a_i a_i^{\ast} \otimes b_i b_i^{\ast}\,:\,i = 1,\ldots,N\}
\]
that perfectly distinguishes $u_1,\ldots,u_m$.
\end{cor}

\noindent
We also note that, without loss of generality, the measurement in this
corollary may be assumed to satisfy the property that $a_i \otimes b_i$ and
$a_j\otimes b_j$ are linearly independent for each choice of $i\not=j$.


\subsection*{Unitary equivalence of realizations of completely positive maps}

The main result of this paper is applied to the question of channel
capacities in Section~\ref{sec:channel}.
It will be helpful in that section to have noted the simple fact below
concerning realizations of completely positive maps.

Let $\trans{\mathcal{X},\mathcal{Y}}$ denote the space of linear mappings
of the form
$\Phi : \lin{\mathcal{X}} \rightarrow \lin{\mathcal{Y}}$.
The {\em Jamio{\l}kowski isomorphism} is the linear mapping
of the form
$J : \trans{\mathcal{X},\mathcal{Y}} \rightarrow
\lin{\mathcal{Y}\otimes\mathcal{X}}$
defined by
\[
J(\Phi) = 
\sum_{i,j} \Phi(E_{i,j}) \otimes E_{i,j}
=
(\Phi \otimes I_{\lin{\mathcal{X}}})(\op{vec}(I_{\mathcal{X}})
\op{vec}(I_{\mathcal{X}})^{\ast}).
\]

\begin{prop}\label{prop:realize}
Suppose that $\Phi\in\trans{\mathcal{X},\mathcal{Y}}$ is completely positive,
and further suppose that $\mathcal{Z}$ is a space and
$A,B\in\lin{\mathcal{X},\mathcal{Y}\otimes\mathcal{Z}}$ are linear mappings
that both realize $\Phi$ in the sense that
\[
\Phi(X) = \tr_{\mathcal{Z}} A X A^{\ast} = \tr_{\mathcal{Z}} B X B^{\ast}
\]
for all $X\in\lin{\mathcal{X}}$.
Then there is a unitary operator $U\in\lin{\mathcal{Z}}$ such that
$A = (I \otimes U) B$.
\end{prop}

\begin{proof}
We have
\begin{align*}
J(\Phi) & = (\Phi \otimes I_{\lin{\mathcal{X}}})(
\op{vec}(I_{\mathcal{X}})\op{vec}(I_{\mathcal{X}})^{\ast})\\
& =
\tr_{\mathcal{Z}} (A \otimes I_{\mathcal{X}})
\op{vec}(I_{\mathcal{X}})\op{vec}(I_{\mathcal{X}})^{\ast}
(A \otimes I_{\mathcal{X}})^{\ast}\\
& = \tr_{\mathcal{Z}} \op{vec}(A)\op{vec}(A)^{\ast},
\end{align*}
and so $\op{vec}(A)\in \mathcal{Y}\otimes\mathcal{Z}\otimes\mathcal{X}$ is a
purification of $J(\Phi)$.
Likewise, $\op{vec}(B)$ is a purification of $J(\Phi)$ as well.
It is well-known that two purifications of a given positive semidefinite
operator are equivalent up to a unitary operator on the space that is
traced out.
In the present situation this implies
\[
\op{vec}(A) = (I_{\mathcal{Y}} \otimes U \otimes I_{\mathcal{X}})
\op{vec}(B) = \op{vec}((I_{\mathcal{Y}}\otimes U)B)
\]
for some unitary operator $U\in\lin{\mathcal{Z}}$.
This is equivalent to $A = (I_{\mathcal{Y}}\otimes U) B$, and so the
proposition is proved.
\end{proof}


\section{Two-way indistinguishability}
\label{sec:indistinguishability}

This section contains a proof of the main result of this paper, which is
stated in the following theorem.

\begin{theorem}
\label{theorem:no-distinguishable-basis}
Let $\mathcal{A} = \mathbb{C}^n$ and $\mathcal{B} = \mathbb{C}^n$ 
for $n\geq 3$.
Then there is no basis of the subspace
$\mathcal{Q}\subseteq\mathcal{A}\otimes\mathcal{B}$ that is perfectly
distinguishable by any LOCC protocol.
\end{theorem}

\noindent
Before giving a formal proof of this theorem, it will be helpful to give a
brief sketch of the proof.
Recall that the operator 
\[
Q = I_{\mathcal{A}\otimes\mathcal{B}} -
\frac{1}{n}\op{vec}(I_{\mathcal{B},\mathcal{A}})
\op{vec}(I_{\mathcal{B},\mathcal{A}})^{\ast}
\]
is the projection onto the subspace $\mathcal{Q}$.
If $\{u_1,\ldots,u_{n^2 - 1}\}$ is a basis of $\mathcal{Q}$ whose elements
are perfectly distinguished by some LOCC protocol, then these elements
are also perfectly distinguished by some rank one separable measurement.
Such a measurement may be written as
\[
\left\{a_i a_i^{\ast} \otimes \overline{b_i} b_i^\t\,:\,
i = 1,\ldots, N\right\}
\]
for $a_1,\ldots,a_N\in\mathcal{A}$ and $b_1,\ldots,b_N\in\mathcal{B}$.
As previously noted, we may assume without loss of generality that
the vectors $a_i\otimes\overline{b_i}$ and $a_j\otimes\overline{b_j}$
are linearly independent for $i\not=j$.
Based on the fact that this measurement perfectly distinguishes the elements
in the chosen basis of $\mathcal{Q}$, we will determine that the basis
\[
\left\{u_1,\ldots,u_{n^2 - 1},
\frac{1}{\sqrt{n}}\op{vec}(I_{\mathcal{B},\mathcal{A}})\right\}
\]
of the entire space $\mathcal{A}\otimes\mathcal{B}$ diagonalizes each of the
operators $Q (a_i a_i^{\ast} \otimes \overline{b_i} b_i^{\t}) Q$
for $1\leq i \leq N$.
Because any two operators that are simultaneously diagonalized by a given
basis must commute, we conclude that
the operators $Q (a_i a_i^{\ast} \otimes \overline{b_i} b_i^{\t}) Q$ and
$Q (a_j a_j^{\ast} \otimes \overline{b_j} b_j^{\t}) Q$ commute for every
choice of $i$ and $j$.
However, based on the properties of the projection $Q$ it can be shown
that there must be a choice of $i$ and $j$ for which
$Q (a_i a_i^{\ast} \otimes \overline{b_i} b_i^{\t}) Q$ and
$Q (a_j a_j^{\ast} \otimes \overline{b_j} b_j^{\t}) Q$ do not commute.
This is a contradiction that stems from the assumption that
$\{u_1,\ldots,u_{n^2 - 1}\}$ is an LOCC distinguishable basis of $\mathcal{Q}$,
and so we conclude that such a basis does not exist.

We now give a more formal proof, beginning with a lemma that proves that there
must exist choices of $i$ and $j$ for which the operators
$Q (a_i a_i^{\ast} \otimes \overline{b_i} b_i^{\t}) Q$
and $Q (a_j a_j^{\ast} \otimes \overline{b_j} b_j^{\t}) Q$ do not commute.

\begin{lemma}\label{lemma:commutation}
Suppose that
\[
\left\{a_i a_i^{\ast} \otimes \overline{b_i} b_i^\t\,:\,
i = 1,\ldots, N\right\}
\]
is a rank one separable measurement such that
$a_i \otimes \overline{b_i}$ and $a_j\otimes \overline{b_j}$ are linearly
independent for all $i\not=j$.
Then there exists a choice of $i$ and $j$ such that the operators
\[
Q (a_i a_i^{\ast} \otimes \overline{b_i} b_i^{\t}) Q
\quad\text{and}\quad
Q (a_j a_j^{\ast} \otimes \overline{b_j} b_j^{\t}) Q
\]
do not commute.
\end{lemma}

\begin{proof}
First note that as
$\left\{a_i a_i^{\ast}\otimes\overline{b_i} b_i^\t\,:\,i=1,\ldots, N\right\}$
describes a measurement, we have
\[
\sum_{i=1}^N a_i a_i^{\ast} \otimes \overline{b_i} b_i^\t = 
I_{\mathcal{A}\otimes\mathcal{B}}.
\]
It follows that 
\[
\op{vec}(I_{\mathcal{B},\mathcal{A}}) = 
\left(\sum_{i=1}^N a_i a_i^{\ast} \otimes \overline{b_i} b_i^\t\right)
\op{vec}(I_{\mathcal{B},\mathcal{A}})
= \op{vec}\left(
\sum_{i=1}^N a_i a_i^{\ast} b_i b_i^\ast\right)
= \op{vec}\left(
\sum_{i=1}^N \ip{a_i}{b_i} a_i b_i^\ast\right)
\]
and therefore
\[
\sum_{i=1}^N \ip{a_i}{b_i} a_i b_i^\ast = I_{\mathcal{B},\mathcal{A}}.
\]
Taking the trace of both sides yields
\[
\sum_{i=1}^N \abs{\ip{a_i}{b_i}}^2 = n.
\]

Now, let
\[
\alpha_{i,j} = (a_i^{\ast} \otimes b_i^{\t}) Q (a_j \otimes \overline{b_j})
\]
for all $i$, $j$.
It will be proved that there exists a choice of $i\not=j$ such that
$\alpha_{i,j}\not=0$.
In order to prove this, assume toward contradiction that
$\alpha_{i,j}=0$ for every pair $i\not=j$.
As
\[
\alpha_{i,j} = (a_i^{\ast} \otimes b_i^{\t}) Q (a_j \otimes \overline{b_j})
= \ip{a_i}{a_j}\ip{b_j}{b_i} - \frac{1}{n}\ip{a_i}{b_i}\ip{b_j}{a_j},
\]
we have that
\[
\ip{a_i}{a_j}\ip{b_j}{b_i} = \frac{1}{n}\ip{a_i}{b_i}\ip{b_j}{a_j}
\]
for all choices of $i\not=j$.
Because $\sum_i\abs{\ip{a_i}{b_i}}^2 = n > 0$, we may choose some
value of $i$ for which $\ip{a_i}{b_i} \not=0$. 
We then have
\begin{multline*}
\ip{a_i}{b_i}  = 
a_i^{\ast}\left(\sum_j \ip{a_j}{b_j}a_j b_j^{\ast}\right)b_i
= \sum_j \ip{a_j}{b_j} \ip{a_i}{a_j} \ip{b_j}{b_i}\\
= \sum_{j\not=i} \ip{a_j}{b_j} \ip{a_i}{a_j} \ip{b_j}{b_i}
+ \ip{a_i}{b_i} \norm{a_i}^2 \norm{b_i}^2\\
= \frac{1}{n} \sum_{j\not=i}\ip{a_j}{b_j} \ip{a_i}{b_i} \ip{b_j}{a_j} 
+ \ip{a_i}{b_i} \norm{a_i}^2 \norm{b_i}^2\\
= \left( \frac{1}{n}\sum_j \abs{\ip{a_j}{b_j}}^2 - \frac{1}{n}
\abs{\ip{a_i}{b_i}}^2 + \norm{a_i}^2 \norm{b_i}^2\right)\ip{a_i}{b_i}\\
= 
\left( 1 - \frac{1}{n}\abs{\ip{a_i}{b_i}}^2 + 
\norm{a_i}^2 \norm{b_i}^2\right)\ip{a_i}{b_i}.
\end{multline*}
As $\ip{a_i}{b_i}\not=0$ this implies
\[
\frac{1}{n}\abs{\ip{a_i}{b_i}}^2 = \norm{a_i}^2 \norm{b_i}^2.
\]
But then by the Cauchy-Schwarz Inequality we have
\[
\abs{\ip{a_i}{b_i}}^2 \leq 
\norm{a_i}^2 \norm{b_i}^2 = \frac{1}{n}\abs{\ip{a_i}{b_i}}^2,
\]
which implies $\abs{\ip{a_i}{b_i}}^2 = 0$.
This contradicts the fact that $i$ was chosen so that $\ip{a_i}{b_i}\not=0$,
and so it has been proved that $\alpha_{i,j} \not= 0$ for some
choice of $i\not=j$.
Fix such a choice for the remainder of the proof.

Next, let us prove that the two vectors $Q(a_i \otimes \overline{b_i})$ and
$Q(a_j \otimes \overline{b_j})$ are linearly independent.
To this end let $\beta$ and $\gamma$ be scalars that satisfy
\[
\beta\,Q(a_i \otimes \overline{b_i}) +
\gamma\,Q(a_j \otimes \overline{b_j}) = 0.
\]
This implies
\[
\beta\, a_i \otimes \overline{b_i} + \gamma\, a_j \otimes \overline{b_j}
= \frac{1}{n}\left(\beta \ip{b_i}{a_i} + \gamma \ip{b_j}{a_j}\right)
\op{vec}(I_{\mathcal{B},\mathcal{A}}),
\]
or equivalently
\[
\beta\, a_i b_i^{\ast} + \gamma\, a_j b_j^{\ast}
= \frac{1}{n}\left(
\beta \ip{b_i}{a_i} + \gamma \ip{b_j}{a_j}\right)I_{\mathcal{B},\mathcal{A}}.
\]
The left hand side of this equation has rank at most 2.
Because we are assuming that $n\geq 3$ this means that the right hand side
must be 0, for otherwise it would have rank $n \geq 3$.
Thus $\beta\, a_i b_i^{\ast} + \gamma\, a_j b_j^{\ast} = 0$, which is
equivalent to
$\beta\, a_i \otimes \overline{b_i} + \gamma\, a_j \otimes \overline{b_j} = 0$.
As $a_i \otimes \overline{b_i}$ and $a_j \otimes \overline{b_j}$ are
necessarily linearly independent, however, this implies that $\beta=\gamma=0$.
Consequently $Q(a_i \otimes \overline{b_i})$ and
$Q(a_j \otimes \overline{b_j})$ are linearly independent

To complete the proof, we must show that the two operators
\[
Q (a_i a_i^{\ast} \otimes \overline{b_i} b_i^{\t}) Q
(a_j a_j^{\ast} \otimes \overline{b_j} b_j^{\t}) Q
= \alpha_{i,j} Q(a_i \otimes \overline{b_i}) (a_j^{\ast} \otimes b_j^{\t})Q
\]
and
\[
Q (a_j a_j^{\ast} \otimes \overline{b_j} b_j^{\t}) Q
(a_i a_i^{\ast} \otimes \overline{b_i} b_i^{\t}) Q
= \overline{\alpha_{i,j}} Q(a_j \otimes \overline{b_j}) 
(a_i^{\ast} \otimes b_i^{\t})Q
\]
are not equal.
Because $\alpha_{i,j} \not= 0$ and the vectors $Q(a_i\otimes \overline{b_i})$
and $Q(a_j\otimes \overline{b_j})$ are nonzero (as they are linearly
independent), neither of these operators is 0.
The images of the two operators are therefore the spaces spanned by the
vectors $Q(a_i\otimes \overline{b_i})$ and $Q(a_j\otimes \overline{b_j})$,
respectively.
The fact that the two operators are not equal therefore follows from the
linear independence of 
$Q(a_i\otimes \overline{b_i})$ and $Q(a_j\otimes \overline{b_j})$.
\end{proof}

\begin{proof}[Proof of Theorem~\ref{theorem:no-distinguishable-basis}]
The proof is by contradiction.
To this end, assume
$\{u_1,\ldots,u_m\}\subset \mathcal{A}\otimes\mathcal{B}$, $m = n^2 -1$, is an
orthonormal basis of $\mathcal{Q}$ whose elements are perfectly distinguished
by some LOCC protocol.
Then there exists a rank one separable measurement
\[
\left\{a_i a_i^{\ast} \otimes \overline{b_i} b_i^\t\,:\,
i = 1,\ldots, N\right\}
\]
for which $a_i \otimes \overline{b_i}$ and $a_j \otimes \overline{b_j}$ are
linearly independent for all $i\not=j$,
together with a partition $S_1\cup \cdots \cup S_m = \{1,\ldots,N\}$,
$S_k\cap S_l = \varnothing$ for $k\not=l$, such that 
\[
u_k^{\ast} \left(\sum_{i\in S_l}
a_i a_i^{\ast} \otimes \overline{b_i} b_i^\t\right) u_k = \delta_{kl}
\]
for all $1\leq k,l\leq m$.

Now, as
\[
u_k^{\ast} \left(a_i a_i^{\ast} \otimes \overline{b_i}b_i^{\t}\right) u_k
= \abs{\ip{u_k}{a_i\otimes \overline{b_i}}}^2,
\]
it follows that $u_k$ and $a_i\otimes\overline{b_i}$ are orthogonal whenever
$i\not\in S_k$.
Consequently, it holds that
\[
u_k^{\ast} 
\left(a_i a_i^{\ast} \otimes \overline{b_i}b_i^{\t}\right) u_l = 0
\]
for $k\not=l$ given that $S_k$ and $S_l$ are disjoint.
The projection $Q$ acts trivially on each of the vectors
$u_1,\ldots,u_m$, and thus
\[
u_k^{\ast} Q
\left(a_i a_i^{\ast} \otimes \overline{b_i}b_i^{\t}\right) Q u_l = 0
\]
for $k\not=l$.
Letting $v = \frac{1}{\sqrt{n}}\op{vec}(I_{\mathcal{B},\mathcal{A}})$
we obviously have $Q v = 0$, and thus
\[
u_k^{\ast} Q \left(a_i a_i^{\ast} \otimes \overline{b_i}b_i^{\t}\right) Q v =
v^{\ast} Q \left(a_i a_i^{\ast} \otimes \overline{b_i}b_i^{\t}\right) Q u_k =
0
\]
for each choice of $k$ as well.
Thus, it has been shown that the orthonormal basis
$\{u_1,\ldots,u_m,v\}$ of $\mathcal{A}\otimes\mathcal{B}$ diagonalizes
each of the operators
$Q \left(a_i a_i^{\ast} \otimes \overline{b_i}b_i^{\t}\right) Q$,
for $1\leq i \leq N$.
As these operators are all simultaneously diagonalized by a common
orthonormal basis, they must therefore commute.
However, by Lemma~\ref{lemma:commutation} this is not the case---for
at least one choice of $i\not=j$ it holds that
$Q \left(a_i a_i^{\ast} \otimes \overline{b_i}b_i^{\t}\right) Q$
and $Q \left(a_j a_j^{\ast} \otimes \overline{b_j}b_j^{\t}\right) Q$
do not commute.
As a contradiction has been reached, this completes the proof of the theorem.
\end{proof}


\subsection*{Impossibility for pairs of qubits}

It should be noted that the assumption $n\geq 3$ in
Theorem~\ref{theorem:no-distinguishable-basis} is necessary.
Indeed, every subspace of a tensor product space
$\mathcal{A}\otimes\mathcal{B}$ where $\mathcal{A} = \mathbb{C}^2$ and
$\mathcal{B} = \mathbb{C}^2$ has a perfectly distinguishable basis.
To see this, let $\mathcal{V}$ be a subspace of $\mathcal{A}\otimes\mathcal{B}$
and let $m = \op{dim}(\mathcal{V})$.
There is nothing to prove for $m = 0$ or $m=1$, the claim for $m = 2$ follows
from Walgate, et al.~\cite{WalgateS+00}, and is trivial for $m = 4$.
In the remaining case $m = 3$, it must be that $\mathcal{V}$ is the
orthogonal complement of some unit vector $u\in\mathcal{A}\otimes\mathcal{B}$.
By considering the Schmidt decomposition of $u$, it is
straightforward to find two product states $a_1\otimes b_1$ and
$a_2\otimes b_2$ so that the set $\{u, a_1\otimes b_1, a_2\otimes b_2\}$ is
orthonormal.
Letting $v$ be any vector orthogonal to the span of
$\{u, a_1\otimes b_1, a_2\otimes b_2\}$, we have that
$\{v, a_1\otimes b_1, a_2\otimes b_2\}$ is an orthonormal basis of
$\mathcal{V}$.
Walgate and Hardy \cite{WalgateH02} have shown that any such set is
perfectly distinguishable given that at least two members of the set
are product states.


\section{Channels with suboptimal classical corrected capacity}
\label{sec:channel}

Hayden and King~\cite{HaydenK04} considered the classical capacity of quantum
channels when the receiver has the capability to measure the channel's
environment and to use the classical result of this measurement when measuring
the output of the channel.
In this section we give examples of channels that have suboptimal capacity
with respect to this definition.
In fact, the capacity of the channels remains suboptimal even when two-way
communication is allowed between the receiver and the environment.

As our aim is to only prove the existence of channels with suboptimal classical
corrected capacity rather than proving quantitative bounds on this capacity,
we will use the following qualitative definition that does not refer to any
specific measure of capacity.
An admissible (i.e., completely positive and trace-preserving) mapping
$\Phi\in\trans{\mathcal{X},\mathcal{A}}$ is said to have {\em optimal
two-way classical corrected capacity} if the following holds.
\begin{mylist}{\parindent}
\item[1.]
There exists a space $\mathcal{B}$ and a unitary embedding
$U\in\lin{\mathcal{X},\mathcal{A}\otimes\mathcal{B}}$ such that
\[
\Phi(X) = \tr_{\mathcal{B}} U X U^{\ast}
\]
for all $X\in\lin{\mathcal{X}}$, and

\item[2.]
there exists an orthonormal basis $\{x_1,\ldots,x_n\}$ of $\mathcal{X}$ such
that the set
\[
Ux_1,\ldots,Ux_n \in \mathcal{A}\otimes\mathcal{B}
\]
is perfectly distinguishable by some LOCC protocol.
\end{mylist}

\noindent
Note that by Proposition~\ref{prop:realize}, a given mapping $\Phi$ fails to
have optimal two-way classical corrected capacity if item 2 above fails to
hold for even a single choice of $U$.
This is because any other choice is equivalent up to a unitary operator
on $\mathcal{B}$, which can simply be absorbed into the LOCC protocol.

The admissible maps that fail to satisfy the above definition are based
on the subspaces considered in the previous section.
Let $n\geq 3$, let $\mathcal{X} = \mathbb{C}^{n^2 - 1}$, and let
$\mathcal{A} = \mathcal{B} = \mathbb{C}^n$.
Choose $u_1,\ldots,u_{n^2 - 1}\in\mathcal{A}\otimes\mathcal{B}$
to be an arbitrary orthonormal basis for the subspace $\mathcal{Q}$
of $\mathcal{A}\otimes\mathcal{B}$.
Define $U\in\lin{\mathcal{X},\mathcal{A}\otimes\mathcal{B}}$
as
\[
U = \sum_{i=1}^{n^2 - 1} u_i e_i^{\ast}.
\]
Obviously $U$ is a unitary embedding, so the mapping
$\Phi\in\trans{\mathcal{X},\mathcal{A}}$ defined by
$\Phi(X) = \tr_{\mathcal{B}} U X U^{\ast}$
for all $X\in\lin{\mathcal{X}}$ is admissible.

\begin{cor}
The mapping $\Phi$ does not have optimal two-way classical corrected capacity.
\end{cor}

\begin{proof}
If $\Phi$ were to have optimal two-way classical corrected capacity, there
would be a choice of an orthonormal basis
$\{x_1,\ldots,x_{n^2 - 1}\}$ of $\mathcal{X}$ such that
$Ux_1,\ldots,Ux_{n^2 - 1}\in \mathcal{A}\otimes\mathcal{B}$
is perfectly distinguishable by an LOCC protocol.
As any such set is necessarily an orthonormal basis of $\mathcal{Q}$, this
cannot be by Theorem~\ref{theorem:no-distinguishable-basis}.
\end{proof}

Although the notion of correctable versus uncorrectable channels does
not require that the input and output spaces have the same dimension,
it is of course simple to adjust such an example to give a channel where
this constraint is satisfied by viewing that the receiver's space
$\mathcal{A}$ is embedded in $\mathcal{X}$.
One may therefore view the example above for $n=3$ as giving a
three-qubit channel having suboptimal two-way classical corrected capacity.


\section{Conclusion}
\label{sec:conclusion}

It has been proved that there exist subspaces of bipartite tensor product
spaces that have no bases that can be perfectly distinguished by LOCC
protocols, and this fact has been used to construct admissible mappings
having suboptimal two-way classical corrected capacity.
There are several interesting unanswered questions relating to these
results, including the following.

\begin{mylist}{\parindent}
\item[1.]
What is the smallest dimension required for a subspace to have no
bases perfectly distinguishable by LOCC protocols?
(The smallest dimension achieved in the present paper is 8.)

\item[2.]
Do there exist subspaces of $\mathcal{A}\otimes\mathcal{B}$ having
no perfectly distinguishable bases when $\op{dim}(\mathcal{A}) = 2$?
As demonstrated in Section~\ref{sec:indistinguishability} this necessarily
requires $\op{dim}(\mathcal{B})\geq 3$.

\item[3.]
Quantitative bounds on the probability with which bases of the subspaces in
question can be distinguished by LOCC protocols were not considered in this
paper, and nor were specific bounds on classical corrected capacities of
the associated channels.
What can be proved about such bounds?

\end{mylist}


\subsection*{Acknowledgments}

I thank Somshubhro Bandyopadhyay, Mehdi Mhalla, and Jonathan Walgate
for several helpful discussions and suggestions.
This research was supported by Canada's NSERC, the Canada Research Chairs
program, and the Canadian Institute for Advanced Research (CIAR).


\bibliographystyle{amsplain}


\providecommand{\bysame}{\leavevmode\hbox to3em{\hrulefill}\thinspace}
\providecommand{\MR}{\relax\ifhmode\unskip\space\fi MR }
\providecommand{\MRhref}[2]{%
  \href{http://www.ams.org/mathscinet-getitem?mr=#1}{#2}
}
\providecommand{\href}[2]{#2}


\end{document}